\documentclass[a4paper,twocolumn,superscriptaddress,11pt, accepted=2018-07-23]{quantumarticle}
\usepackage{braket}
\usepackage{graphicx}
\usepackage{dsfont}
\usepackage{amsmath}
\usepackage{hyperref}
\usepackage{hyphenat}

\usepackage[numbers, sort&compress]{natbib}

\graphicspath{ {images/} }

\newcommand{\Tr}[1]{\text{Tr}[#1]}
\newcommand{\figref}[1]{Fig.\ref{#1}}
\renewcommand{\eqref}[1]{Eq.(\ref{#1})}

\begin{document}

\title{In situ upgrade of quantum simulators to universal computers}
\date{2018-07-23}

\author{Benjamin Dive}
\affiliation{Department of Physics, Imperial College, SW7 2AZ London, UK}
\orcid{0000-0002-7523-8983}

\author{Alexander Pitchford}
\affiliation{Institute of Mathematics, Physics and Computer Science, Aberystwyth University, SY23 3FL Aberystwyth, UK}
\orcid{0000-0002-4717-2921}

\author{Florian Mintert}
\affiliation{Department of Physics, Imperial College, SW7 2AZ London, UK}
\orcid{0000-0001-8213-4368}

\author{Daniel Burgarth}
\affiliation{Institute of Mathematics, Physics and Computer Science, Aberystwyth University, SY23 3FL Aberystwyth, UK}
\orcid{0000-0003-4063-1264}

\nohyphens{\maketitle}

\begin{abstract}
Quantum simulators, machines that can replicate the dynamics of quantum systems, are being built as useful devices and are seen as a stepping stone to universal quantum computers. A key difference between the two is that computers have the ability to perform the logic gates that make up algorithms. We propose a method for learning how to construct these gates efficiently by using the simulator to perform optimal control on itself. This bypasses two major problems of purely classical approaches to the control problem: the need to have an accurate model of the system, and a classical computer more powerful than the quantum one to carry out the required simulations. Strong evidence that the scheme scales polynomially in the number of qubits, for systems of up to $\mathbf{9}$ qubits with Ising interactions, is presented from numerical simulations carried out in different topologies. This suggests that this \textit{in situ} approach is a practical way of upgrading quantum simulators to computers.\end{abstract}

Recent and ongoing work on building large quantum systems is leading to simulators that are able to model physical phenomena, allowing questions about the underlying science to be answered \cite{Cirac2012, Johnson2014, Georgescu2014}. These machines contain a register of quantum particles, typically two level quantum systems (qubits) storing quantum information. The presence of interactions between these leads to dynamics that, by varying control parameters in the system Hamiltonian, can replicate the quantum behaviour of systems of interest. This, however, is less general than a quantum computer which is able is to perform a universal set of logic gates on the qubits \cite{DiVincenzo2000}.

\begin{figure}
\centering
\includegraphics[trim = 8cm 1cm 8cm 0cm, width=0.84\columnwidth]{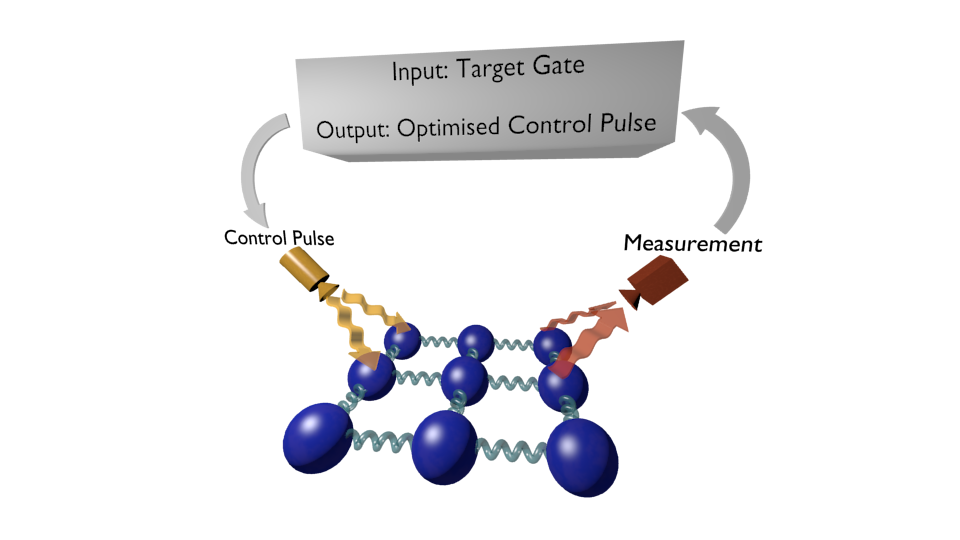}
\caption{A classical computer finds a control pulse which enables a quantum simulator to perform logic gates. It does this in an iterative process by applying a control pulse to the simulator and then improving it based on the result of measurements.}
\label{fig:full_schematic}
\end{figure}

Provided some control parameters can be varied in time, it is in principle possible to do an arbitrary gate on a quantum many-body system such as a quantum simulator \cite{Lloyd1995, Dodd2002, Burgarth2009}. Finding the right time-dependency however relies almost exclusively on numerical methods, especially when physical constraints on the control fields are taken into account \cite{Machnes2011, Glaser2015}. These methods require a very precise knowledge of the parameters of a system, a daunting task for a machine with a huge number of degrees of freedom. Furthermore, they are intractable on a classical computer if the quantum simulator we want to solve the problem for is large enough to do something beyond the capabilities of classical computers. These two difficulties provide a major obstacle in using quantum simulators to perform arbitrary computation.

We circumvent these problems by showing how well-known existing numerical methods can be translated to run \emph{in situ} on the quantum simulator itself, as illustrated in \figref{fig:full_schematic}. A mix of analytical and numerical results point towards this being a scalable, bootstrapping scheme for performing a universal quantum computation, needing resources that grow only polynomially with the number of qubits. 

The core principle is to use the system at hand to implement and improve control pulses until they perform the desired gate on it. Such an adaptive approach to finding controls is naturally used in laboratory work and has been studied as a fine-tuning tool in small systems \cite{Egger2014, Kelly2014}, a way of controlling quantum chemistry with light \cite{Rabitz2016}, a method for stochastic optimisation \cite{Ferrie2015}, and a way of correcting parameter drift \cite{Kelly2016}. Using simulators as oracles for reaching quantum states has also been explored \cite{Li2016}, as has the potential for a quantum speed up in general optimisation problems \cite{Rebentrost2016}. In this paper we propose a way to transform this general approach into a large scale, constructive method that provides a new avenue to control many-body quantum systems. Independently, and concurrently to the writing of this paper, a similar approach as ours was developed and tested experimentally for the different problem of quantum state preparation \cite{Lu2017}. 

\subsection*{The scheme}
The model we consider is a quantum simulator that has the underlying ability to be a universal quantum computer (one that can run an arbitrary algorithm), and the task is to learn how to use it as such. For this reason, we take the simulator as consisting of $n$ qubits with some interactions between them such that they form a fully connected graph where every qubit is (directly or indirectly) interacting with every other qubit. Furthermore we require that the timescale associated with this interaction is much shorter than the decoherence time in order for significant entanglement to be built up. In addition to this we need the ability to perform the following operations on each qubit individually: preparation in a complete basis set of states, fast rotations by applying strong Hamiltonians, and measurement in a complete basis set.

Such a system can be described by the Hamiltonian
\begin{equation}
H = \sum_i\left(f_x^i(t) \sigma_x^i + f_y^i(t) \sigma_y^i\right) + \sum_{<i,j>} H^{i, j},
\label{eq:systemHam}
\end{equation}
where the first sum is over all qubits and the time-dependent control functions $f(t)$ are to be determined. The second sum is over all connected qubits and $H^{i, j}$ is the interaction between qubit $i$ and $j$. The choice of $\sigma_x$ and $\sigma_y$ for the controls is for convenience, any two Hamiltonians will work, and does not have to be the same for all qubits. As the controls are found \emph{in situ}, it is not necessary to know beforehand the form of the interactions $H^{i, j}$.

These requirements are significant, but much easier than demanding direct control over two qubit operations, and correspond to the state-of-the-art in systems involving trapped ions \cite{Johanning2009, Lanyon2011, Blatt2012}, cold atoms \cite{Bloch2012, Labuhn2015}, NMR \cite{Peng2010, Cai2013, Silva2016} or superconducting circuits \cite{Houck2012, OMalley2015}. In these systems there already exist quantum simulators powerful enough to do simulations, and satisfy our requirements, but are not currently usable as computers as it is not known how to perform logic gates on them \cite{DiVincenzo2000, Johnson2014}. Our numerical results show that the scheme developed here scales well for a range of systems where $H^{i, j}$ is of the Ising type ($\sigma_z\otimes\sigma_z$). As Ising machines are useful for a wide range of quantum simulations and can be built with many different technologies \cite{Lanyon2011, Labuhn2015, McMahon2016}, this is a result with wide ranging applicability.

In the model we consider, the connectedness of the qubits and the ability to do fast single qubit operations guarantees that the two core requirements of our proposed optimisation scheme are satisfied: there exists a universal gate set that can be reached at short times \cite{Dodd2002}, and process tomography can be performed \cite{Poyatos1997}. While other systems satisfy these requirements and the approach detailed here would work, we focus on this model for clarity. As single qubit operations are assumed, the gates that controls are needed for are entangling ones, canonically the controlled-not (C-NOT) gate; these are vastly harder to perform using conventional methods and typically have much lower fidelities. 

\begin{figure}
\centering
\includegraphics[trim = 1.5cm 0.5cm 1.5cm 0cm, width=.85\columnwidth]{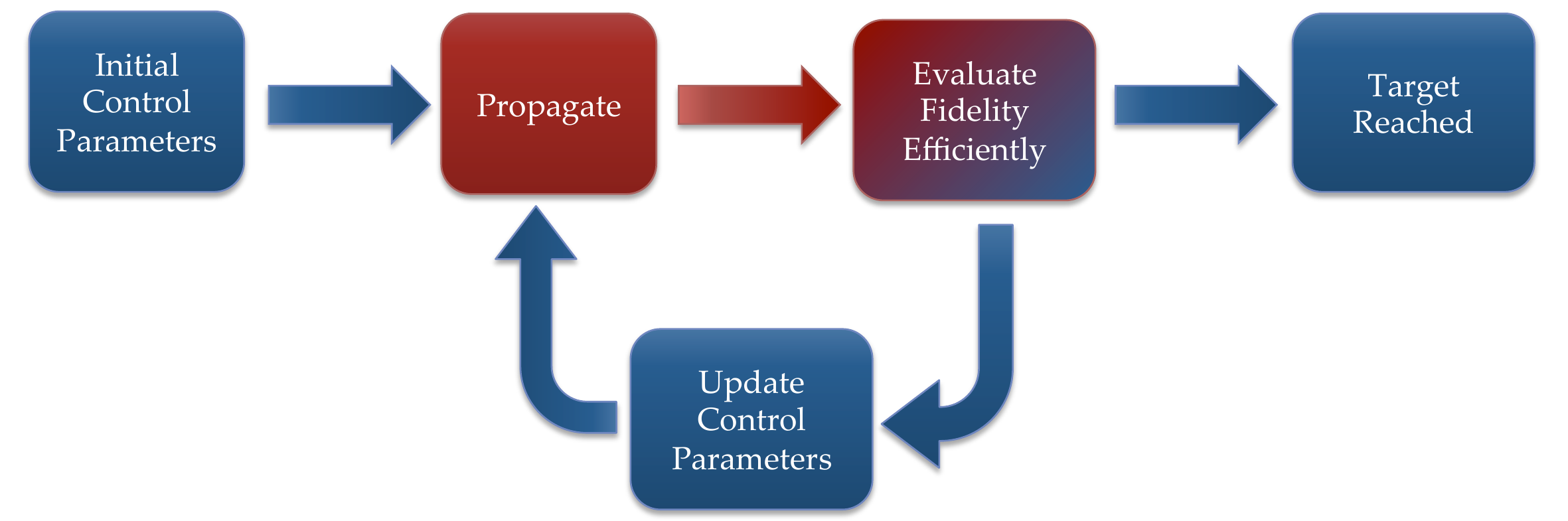}
\caption{Outline of the process used in optimal control, the two red steps in the middle are done \emph{in situ} in our scheme while the others are classical processing. The starting point is an initial set of controls that parametrise the strength of the control Hamiltonians over the gate duration, in our examples these are generated randomly. The evolution of the system with these parameters is then calculated. On a classical computer this requires solving the time-dependent Schr\"odinger equation numerically for a model of the system, while in our scheme this is simply implementing the controls on the simulator. Evaluating the gate fidelity in the classical case is straightforward but, when done in situ, requires some form of tomography to measure it. We derived a tight bound for this gate fidelity in \eqref{eq:fidbound} that can be measured efficiently. If this fidelity is above a threshold, the process terminates successfully, otherwise the control parameters are updated based on the results of the latest and previous runs, and the process repeats. Such an approach can be used in a wide range of contexts, such as to perform quantum logic using random walks \cite{Lahini2018}.}
\label{fig:flowchart}
\end{figure}

The steps for finding such a gate in the \emph{in situ} scheme are outlined in \figref{fig:flowchart}. These are very general and almost the same as in classical numerical optimisation: a guess for the optimal values is generated, these are fed into a function that computes their effect, the distance between this and the desired outcome is calculated, and this generates another guess for the optimal values. This is iterated until the values get close enough to the goal, or the process terminates unsuccessfully after some timeout condition is reached. The difficulty with doing this for a quantum simulator of the type discussed above is in computing what unitary is produced by a given choice of control parameters; this requires both an accurate model of a high dimensional system and an exponentially large classical computer to solve it. Neither of those things can be done for a quantum system large enough to be an interesting quantum computer.

We eliminate these twin difficulties by using the quantum simulator to compute the effects of the control pulse on itself. This works because the simulator with a trial set of controls is guaranteed to be an accurate model of itself with those controls. The propagation step is therefore done \emph{in situ}, but the method by which the control parameters are updated remains purely classical. This is because the information extracted from the quantum simulator (the gate fidelity) and the parametrisation of the control pulses are purely classical. An upshot of this is that the myriad of different methods to do numerical optimisation that have already been developed and work for quantum systems can be used in this protocol directly.

In order to use the quantum simulator as a universal computer, this optimisation procedure needs to be repeated for a universal set of gates. As single qubit operations are assumed, it is sufficient to find a complete set of C-NOT gates. The minimum number of these gates, such that every quantum circuit can be implemented, is $n-1$. In practice we expect it to be more efficient, and produce shorter circuits, to find the $\tfrac{1}{2} n (n-1)$ C-NOT gates that act between every pair of qubits. As is shown in the numerical section, this is readily achieved even for pairs of qubits that are not directly interacting.

The question of whether the scheme works when the system Hamiltonian varies in time uncontrollably is important for an experimental implementation. If this change in time happens slowly compared to the time it takes to find a control pulse for a given gate, then it does not impede the ability of the scheme to find that pulse. However, it may mean that the pulse no longer produces the required dynamics at a later point in time when it is being used as part of an algorithm. In this case, it would be required to eventually rerun the optimisation scheme in order to correct for this drift.

On the other hand, if the stochastic fluctuations in the Hamiltonian are faster than the evolution time required for a gate, the problem is a different one. Repeatedly evolving the system with the same controls (necessary in order to measure the fidelity) will result in the system evolving with a different Hamiltonian on each repeat. This noise in the Hamiltonian thus translates into a lower fidelity being measured. Therefore, as long as the fast fluctuations in the Hamiltonian are small, they are not expected to prevent the scheme from finding a successful pulse but will limit the maximum possible fidelity.

\subsection*{Local gate fidelity}\label{sec:LocalFidelity} The measure used to gauge how close the system evolution is to the desired unitary is typically the gate fidelity \cite{Gilchrist2005}. This is a function between the dynamical map $M$ which describes the evolution of the system under a set of controls (including potential decoherence which acts on the system), and the target unitary $U$. It is defined as $F(M, U) = \bra{\psi} \rho_M \ket{\psi}$ where $\ket{\psi} = U \otimes \mathds{1} \ket{\Omega}$ (with $\ket{\Omega} = \sum_k \ket{kk}$ being a maximally entangled state) is the Choi state of $U$; and $\rho_M = (M \otimes id) \ket{\Omega}\bra{\Omega}$ is the Choi state of $M$. This distance measure is bounded between $0$ and $1$, with the upper limit being reached only when $M(\cdot) = U(\cdot)U^\dagger$. In the case of the system evolution being  unitary simplifies down to $F(V, U) = |\tfrac{1}{d} \Tr{V^\dagger U}|^2$. When the propagation step of \figref{fig:flowchart} is done classically, the whole unitary describing the evolution of the system is calculated as an exponentially large matrix from which the gate fidelity must be calculated.

In the \emph{in situ} scheme this is no longer the case; the only thing which is accessible is a quantum state after it has been evolved by the quantum simulator under some control parameters. The standard method of extracting the gate fidelity is to perform a variant of process tomography, known as certification. This requires preparing the system in a specific state, evolving it, and then performing a set of measurements. The number of different preparation-measurement combinations, $N_{\text{meas}}$, is of order $O(d^2)=O(2^{2n})$, and thus scales exponentially \cite{DaSilva2011}.

However it is possible to do exponentially better for cases of interest where the target gate has a tensor product structure, $U = \bigotimes U_i$ where each $U_i$ is a unitary which acts on a small number of qubits. This would typically be a single C-NOT on one pair of qubits and identity on the rest, $\text{C-NOT}_{1,2} \otimes \mathds{1}_3 \otimes \mathds{1}_4 \otimes ...$, but it could be several simultaneous non-overlapping C-NOTs or even larger gates such as Tofolli. No matter what the exact form is, provided that the target can be decomposed into a tensor products of unitaries, the fidelity over the whole system is bounded by the \emph{local estimator} $F_{LE}$ according to
\begin{equation}
F(M, U) \ge F_{LE}(M,U) = 1 - \sum_i (1 - F(M_i, U_i))
\label{eq:fidbound}
\end{equation}
where $M_i(\rho_i) = M(\rho_i \bigotimes_{j\ne i} \tfrac{1}{d_j} \mathds{1}_j)$. This is the reduced dynamical map acting on subsystem $i$ where the other subsystems have been initialised in the maximally mixed state. This result is proved in the appendix below based on existing approaches \cite{Cramer2010}.

\begin{figure*}
\centering
\includegraphics[scale=0.5]{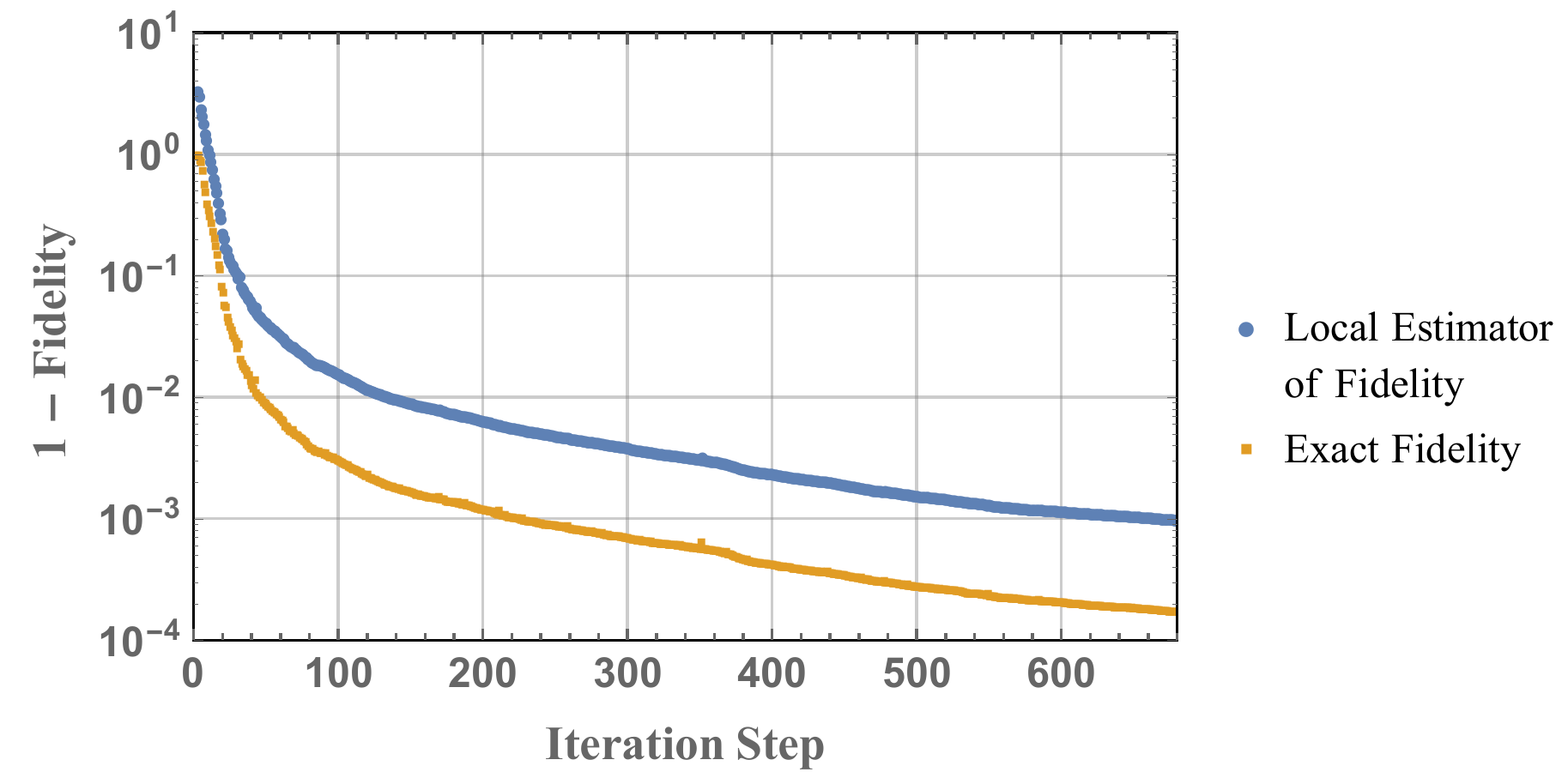}
\includegraphics[scale=0.5]{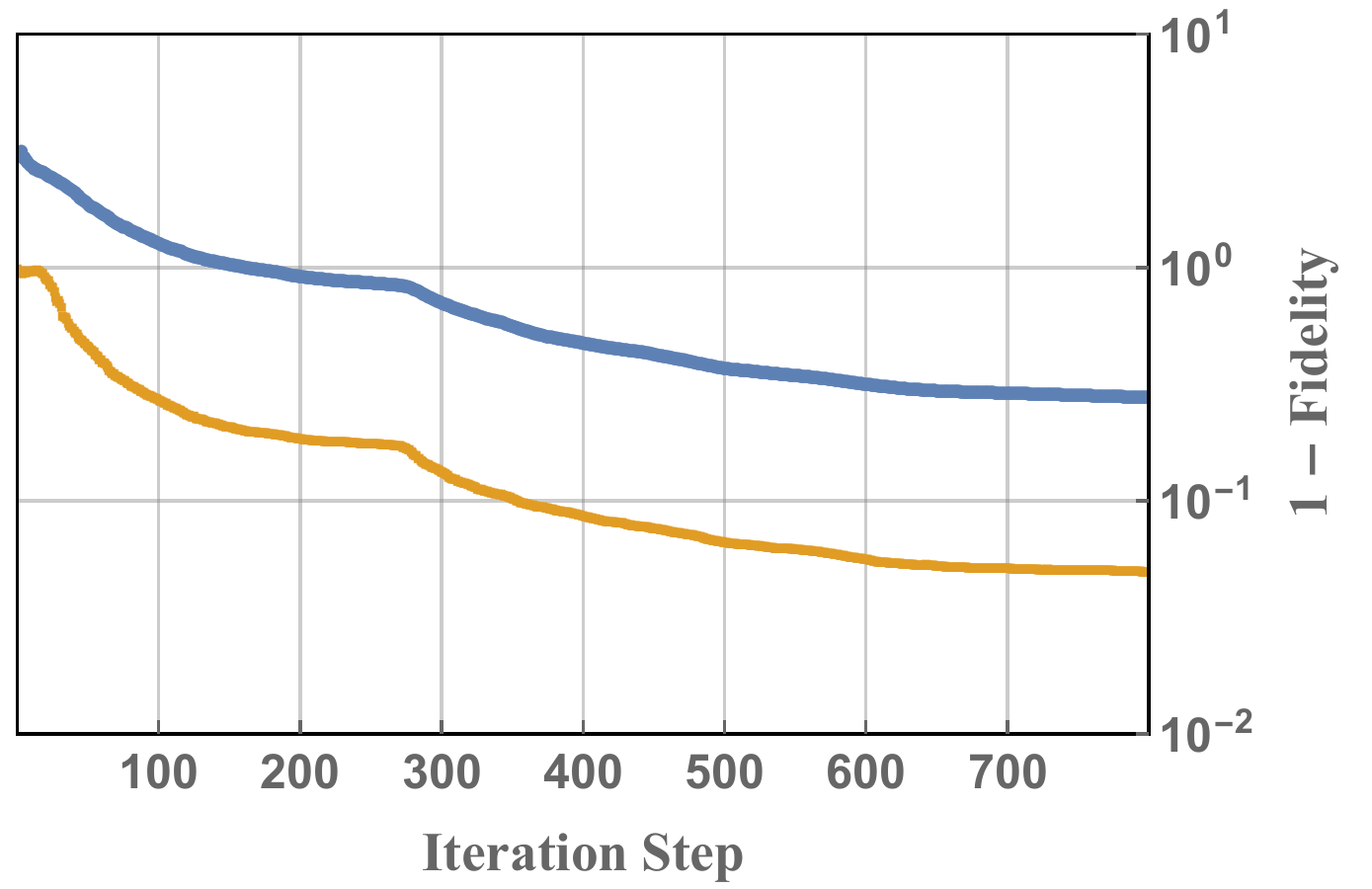}
\\
\small{a) Ising Chain \hspace{0.32\linewidth} b) Heisenberg Chain} 
\caption{ \bf Comparison of the gate fidelity with the local estimator of the fidelity during an optimisation run.
\normalfont
The gate fidelity and its local estimator, \eqref{eq:fidbound}, are plotted as a function of iteration step for one complete run of the \emph{in situ} optimisation scheme. The system is a five-qubit nearest-neighbour chain with the Hamiltonian of \eqref{eq:systemHam}; Ising on the left where $H^{i, j} = \sigma_z \otimes \sigma_z$, and Heisenberg on the right where $H^{i,j} = \sigma_x\otimes\sigma_x + \sigma_y\otimes\sigma_y + \sigma_z\otimes\sigma_z$. The target is a C-NOT gate on the first two qubits and identity on the others. The algorithm minimised the infidelity of the local estimator.
The exact infidelity is plotted at each step for comparison. It is lower in both cases at all iteration steps, and highly correlated with the estimated infidelity, such that minimising the former also minimises the latter almost monotonically and the landscape remains trap free. Furthermore the difference between the two decreases rapidly as the infidelity approaches $0$. In the Heisenberg case the true gate fidelity converges slower than for the Ising chain; this behaviour is closely mapped onto the local estimator. This demonstrates the validity of maximising the local estimator of the fidelity as a proxy for maximising the true gate fidelity.}
\label{fig:FidelityCurves}
\end{figure*}

The advantage of this local estimator to the fidelity is that it only requires certification to be performed over small dimensional subsystems of $1$ or $2$ qubits. As the size of these subsystems does not increase as the system is scaled up, the cost of measuring the fidelity does not increase exponentially with the number of qubits. Applying existing results for certification to each term in \eqref{eq:fidbound} sequentially gives $N_{\text{meas}}=O(\sum_i d_i^2)=O(n)$. As is discussed in the methods section below, it is possible to remove this linear scaling by noting that each term in \eqref{eq:fidbound} can be recovered in parallel. This results in a constant cost, $N_{\text{meas}}=O(\max_i d_i^2)=O(1)$, a vast improvement over the previous exponential scaling, $O(2^{2n})$.

Beyond being efficiently recoverable, this estimator to the fidelity is useful for a number of reasons. It is a lower bound on the gate fidelity, so we are guaranteed that the true fidelity is at least as good. It converges to the exact fidelity in the limit that $F(M, U) \to 1$, this is important as we are most interested in having a measure of how good a gate is when it is close to the target. As can be seen in \figref{fig:FidelityCurves}, it is well behaved numerically and the initial convergence is fast. Increasing the number of qubits would increase the number of terms in \eqref{eq:fidbound} but not their structure, therefore we expect the qualitative features to remain the same as it is scaled up. The minimum value of the local fidelity is $1-n$, while the true gate fidelity cannot go below $0$, so it may be expected that the convergence to $1$ is slower in larger systems as the local fidelity has a larger range to cover.

The scaling behaviour of the local fidelity was investigated by considering a target $U_T = \text{C-NOT} \otimes\mathds{1}\otimes\mathds{1}...$ and comparing the gate fidelity and local fidelity between it and the unitary $U = e^{-i H} U_T$ , where $H$ is a random Hamiltonian generated by a Gaussian distribution and normalised to $||H||_2 = 0.1$. As the number of qubits was increased from $3$ to $14$, the true gate fidelity averaged over different random $H$ stayed the same ($~99.5\%$) while the local fidelity dropped linearly, by less than $0.2\%$ per qubit. This is in accordance with our intuition that the local fidelity behaves similarly for different numbers of qubits, with the principle difference being the linearly increasing number of terms in the sum of \eqref{eq:fidbound}.

\subsection*{Numerical investigation} The local fidelity detailed in the previous section shows that it is possible to estimate the fidelity of a quantum gate efficiently as the size of the system increases, removing one direct barrier from the scalability of the \emph{in situ} optimisation scheme. There are, however, other factors that determine the time the protocol takes which need to be taken into account to assess its scalability. This requires an expression for the total time required to construct a control sequence for a gate in terms of the number of qubits in the system. As this is an optimisation problem that would be done `numerically' on a hybrid classical-quantum computer, analytic expressions could not be obtained. In order to investigate this we conducted simulations of the protocol on a purely classical computer. We explored systems from $3$ to $9$ qubits; memory constraints on the cluster we used made larger systems unfeasible due to the difficulty of evolving (and doing gradient based optimisation of) operators.

The average time needed to find a control sequence for a gate can be expressed as:
\begin{equation}
T_{\text{total}} = T_{\text{run}} N_{\text{runs}} / p_{\text{succ}}
\label{eq:time}
\end{equation}
where $T_{\text{run}}$ is the time it takes to do one run of a control sequence on the quantum simulator, $N_{\text{runs}}$ is the number of sequences that are run on the simulator until the protocol halts, and $p_{\text{succ}}$ is the probability that the protocol halts with a control pulse that reaches the desired fidelity. $T_{\text{run}}$ can be decomposed as $T_{\text{run}} = T_{\text{init}} + T_{\text{gate}} + T_{\text{meas}}$ which is the time to initialise the system, evolve the system under the interaction and control Hamiltonians, and then measure it respectively. $T_{\text{init}}$ and $T_{\text{meas}}$ are determined by the type of system being used; we take them as fixed and independent of the number of qubits. The gate time, on the other hand, is a free parameter that must be decided before starting the \emph{in situ} optimisation.

The total number of runs can be similarly decomposed as
\begin{equation}
N_{\text{runs}} = N_{\text{meas}}\;N_{\text{prec}}\;N_{\text{fids}}\;N_{\text{upds}}.
\label{eq:cost}
\end{equation}
$N_{\text{meas}}$ is the number of times the experiment with the same control pulse, but with different input states and measurement basis, must be repeated in order to measure the gate fidelity once. As the previous section showed, this is $O(1)$ for the local estimator to the fidelity, which is the measure used henceforth. $N_{\text{prec}}$ is the number of times the fidelity must be measured to acquire sufficient statistics such that the fidelity is known to the desired precision. $N_{\text{fids}}$ is the number of different fidelities that need to be measured for the optimisation algorithm to update the control sequence. It is $1$ for gradient-free methods, while for steepest-ascent methods it is $1$ plus the number of gradients (when they are measured by finite difference). $N_{\text{upds}}$ is the number of the times the control sequences must be updated, corresponding to the number of times the scheme goes around the loop of \figref{fig:flowchart}.

The scaling relation of the terms in \eqref{eq:time} depends on the underlying classical algorithm. We used a steepest ascent gradient method similar to Gradient Ascent Pulse Engineering (GRAPE) algorithm \cite{Khaneja2005} which is commonly used for optimising quantum control on classical computers with great success. In this approach, each of the independent Hamiltonians that can be controlled are taken as piecewise-constant with $N_{\text{ts}}$ time-slots of equal widths that span the full gate time $T_{\text{gate}}$. We used analytical gradients in the simulations, for computational efficiency \cite{Machnes2011}, which restricted us to piecewise constant and fixed $T_{\text{gate}}$.

The precision to which the fidelities need to be measured experimentally also needs to be specified. This could be done by either fixing $N_{\text{prec}}$ itself, or by repeatedly measuring the fidelity until the error of the mean is below a specified value. We approximated the latter approach numerically by rounding each local fidelity measurement to some numerical accuracy $A_{\text{num}}$ and calculated the equivalent $N_{\text{prec}}$. The \emph{in situ} scheme therefore requires $T_{\text{gate}}$, $N_{\text{ts}}$ and $A_{\text{num}}$ to be chosen beforehand, as well as a target fidelity $F_{\text{targ}}$, and to know the number of different control Hamiltonians which, multiplied by $N_{\text{ts}}$, gives the total number of controls $N_{\text{ctrl}}$.

In order to simulate this completely numerically we also need to specify exactly what the control Hamiltonians and the constant interaction Hamiltonians are for a given system. This is not the case were this scheme done \emph{in situ} experimentally. In an experiment $T_{\text{gate}}$ minimisation could be included in optimisation objectives as the gradients would be calculated via a finite-difference method. Alternative pulse parametrisation to piecewise constant could also be used, as best suits the experimental setup. All the different parameters mentioned above are summarised in \figref{fig:parametersummary}.

\begin{figure}
\small
\begin{center}
\begin{tabular}{| c | l |}
\hline
Parameter & Description \\
\hline
$T_{\text{gate}}$ & Evolution time for the gate \\
$N_{\text{ts}}$ & Number of timeslots for control pulse \\
$A_{\text{num}}$ & Accuracy of fidelity measurements \\
$F_{\text{targ}}$ & Target fidelity for the desired gate \\
\hline
$N_{\text{runs}}$ & Number of (\#) runs in total\\
$N_{\text{meas}}$ & \# different input-output pairs\\
$N_{\text{prec}}$ & \# repeats for required fidelity accuracy\\
$N_{\text{fids}}$ & \# different fidelities to update controls\\
$N_{\text{upds}}$ & \# control updates needed\\
$p_{\text{succ}}$ & probability of success\\ 
$N_{\text{ctrl}}$ & \# parameters in control pulse\\
\hline
\end{tabular}
\end{center}
\caption{The different parameters defined in the text, summarised here for convenience. The top four are those which need to be fed into the classical optimiser in order for it to run GRAPE \emph{in situ}; other classical protocols could be used, which would require different parameters. The bottom seven are used to quantify the efficiency of the scheme.}
\label{fig:parametersummary}
\end{figure}

We conducted a number simulations of this approach with a Hamiltonian of the type described in \eqref{eq:systemHam} for different number of qubits and interaction topologies. They were completed using the quantum optimal control modules in QuTiP  \cite{Johansson2012, Johansson2013, qutipweb}. These provide methods for optimising a control pulse to some fidelity measure. The GRAPE implementation in QuTiP is described in the documentation, available at \cite{qutipweb}. The code used to perform the numerical simulations is available in an open-source repository \cite{qinsitu2018}. The \emph{local Choi fidelity} measure customisation, and a method for automating locating the $p_{\text{succ}}$ threshold, were developed for this study; they are fully described in the code documentation. The optimal $T_{\text{gate}}$ and $N_{\text{ts}}$ were determined by trialling a range of alternatives. As the result of the optimisation depends on the initial random control amplitudes (uniformly distributed in $[1,1]$), each scenario was repeated multiple times to gain reliable statistics.

A high performance computing cluster was necessary for completing sufficient repetitions of the optimisation simulations of the larger systems in a reasonable time (the 9 qubit optimisations each required around four days of Intel Xeon CPU E5-2670 0 2.90GHz core processing time and were repeated hundreds of times). This is because the processing time required to optimise a pulse scales exponentially with system size due to the need to exponentiate the Hamiltonians in order to compute propagators. This difficulty precisely highlights the need to optimise pulses \emph{in situ} for quantum systems of the size that would perform a useful quantum computation. 

\begin{figure}
\small
\begin{center}
\begin{tabular}{| c | c | c | c | c |}
\hline
Topology & Coupling & $T_{\text{gate}}$ & $N_\text{ts}$ & $N_\text{upds}$ \\
\hline
chain & Ising & $\pi$ & $12$ & $60$  \\
star & Ising & $\pi$ & $12$ & $214$  \\
fully connected & Ising & $12\pi$ & $160$ & $295$  \\
chain & Heisenberg & $16\pi$ & $160$ & $585$ \\
star & Heisenberg & $12\pi$ & $160$ & $1043$  \\
fully connected & Heisenberg & $12\pi$ & $160$ & $881$  \\
\hline
\end{tabular}
\end{center}
\caption{The cost of performing the \emph{in situ} optimisation scheme is investigated for a range of different 5 qubit systems for the Hamiltonian of \eqref{eq:systemHam}. The differences between the systems are their topology (a linear chain with nearest neighbour interactions, a star where all interact with a central qubit only, or fully connected where the interaction strengths are also randomised) and the interaction type. In each case the Hamiltonian used means that $N_{\text{ctrl}}=10$, the target operation is a C-NOT gate on two qubits and identity on the rest, $F_{\text{targ}} = 0.999$, and $p_{\text{succ}} > 0.98$. These simulations where done with full numerical precision. We see that, for five qubits, all six systems find the desired entangling gate, and do so at reasonable experimental cost. This indicates that the approach works for a range of possible quantum simulators.}
\label{fig:topologies}
\end{figure}

We found numerically that, for a range of examples, there exist values of $T_{\text{gate}}$ and $N_{\text{ts}}$ such that the \emph{in situ} scheme converges. \figref{fig:topologies} shows typical values of the most important parameters for a variety of topologies and interaction types. We consistently found that Ising systems were easier to find controls for than Heisenberg systems. In particular, our results suggest that in Heisenberg systems a GRAPE-based algorithm may require a  $T_{\text{gate}}$ that scales exponentially with the number of qubits in order for the optimisation to succeed. We found that this discrepancy also existed in purely classical optimisation techniques.

This suggests that Heisenberg systems are intrinsically harder to solve with optimal control methods than Ising ones, and that this does not depend on whether an \emph{in situ} or classical approach is used. This is consistent with \figref{fig:FidelityCurves} where we compare the local estimator to the fidelity with the true fidelity for a $5$ qubit Heisenberg chain as a function of the iteration step as it is being optimised. The local estimator tracks the true fidelity steadily whether the underlying system is Heisenberg or Ising, the notable difference between the two plots is the plateau in the Heisenberg case. This shows that optimising the system is significantly harder and appears for both the local estimator and the true fidelity. An exponential scaling in the required gate time would make such an approach infeasible for a quantum computer.

Regardless of these numerical difficulties, it can be shown \cite{Dodd2002} that it is possible to do fast entangling gates on such Heisenberg systems by using fast local unitaries and Trotter compositions to decouple the system into simple disconnected components. The problem is therefore with the choice of the particular optimisation algorithm that struggles to find the solution. It may be the case that using a different algorithm inside our \emph{in situ} protocol, such as parametrising the control Hamiltonians as a Fourier series rather than as piecewise-constant, would find control pulses for shorter gate times.

\begin{figure*}[t]
\centering
a)\includegraphics[scale=0.48]{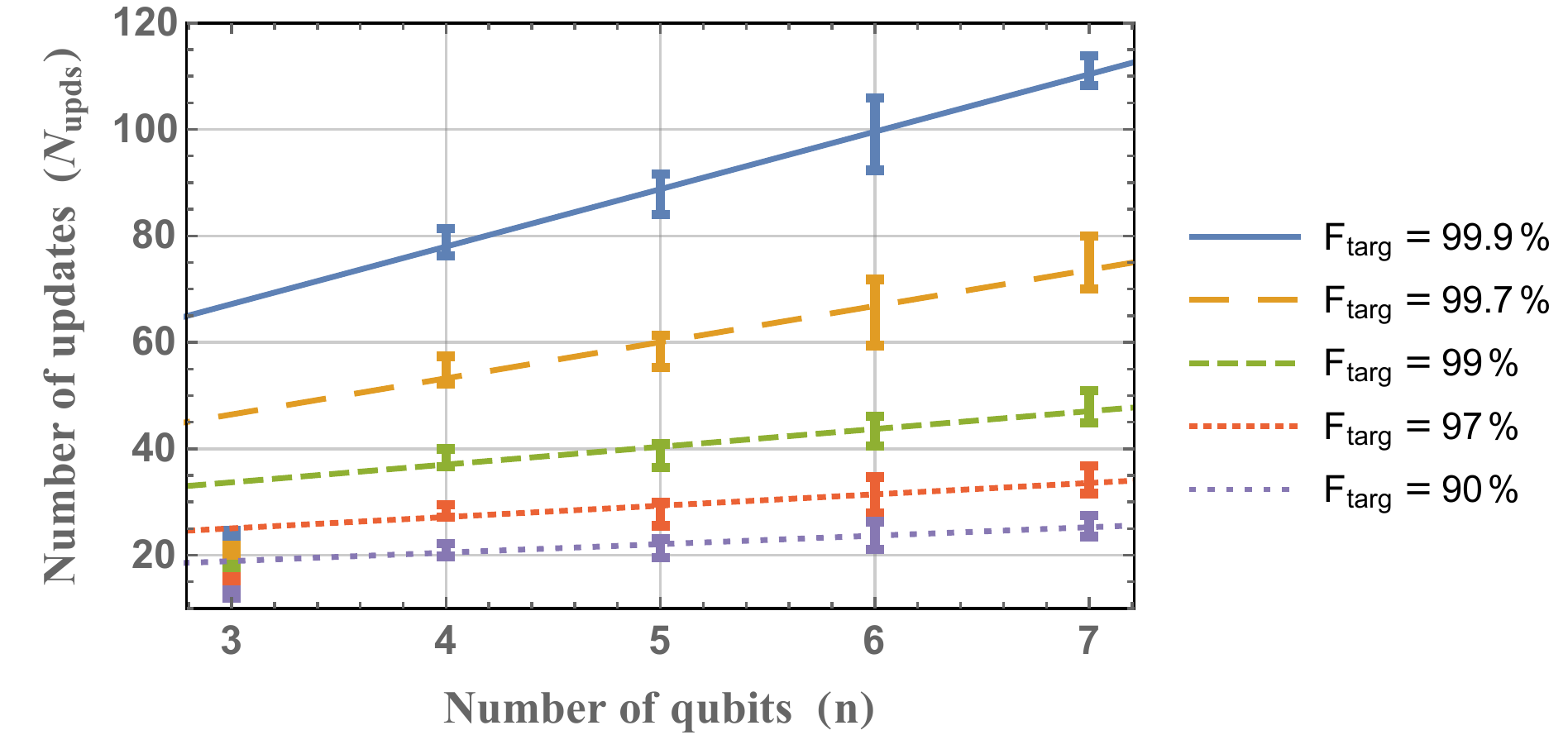}
b)\includegraphics[scale=0.48]{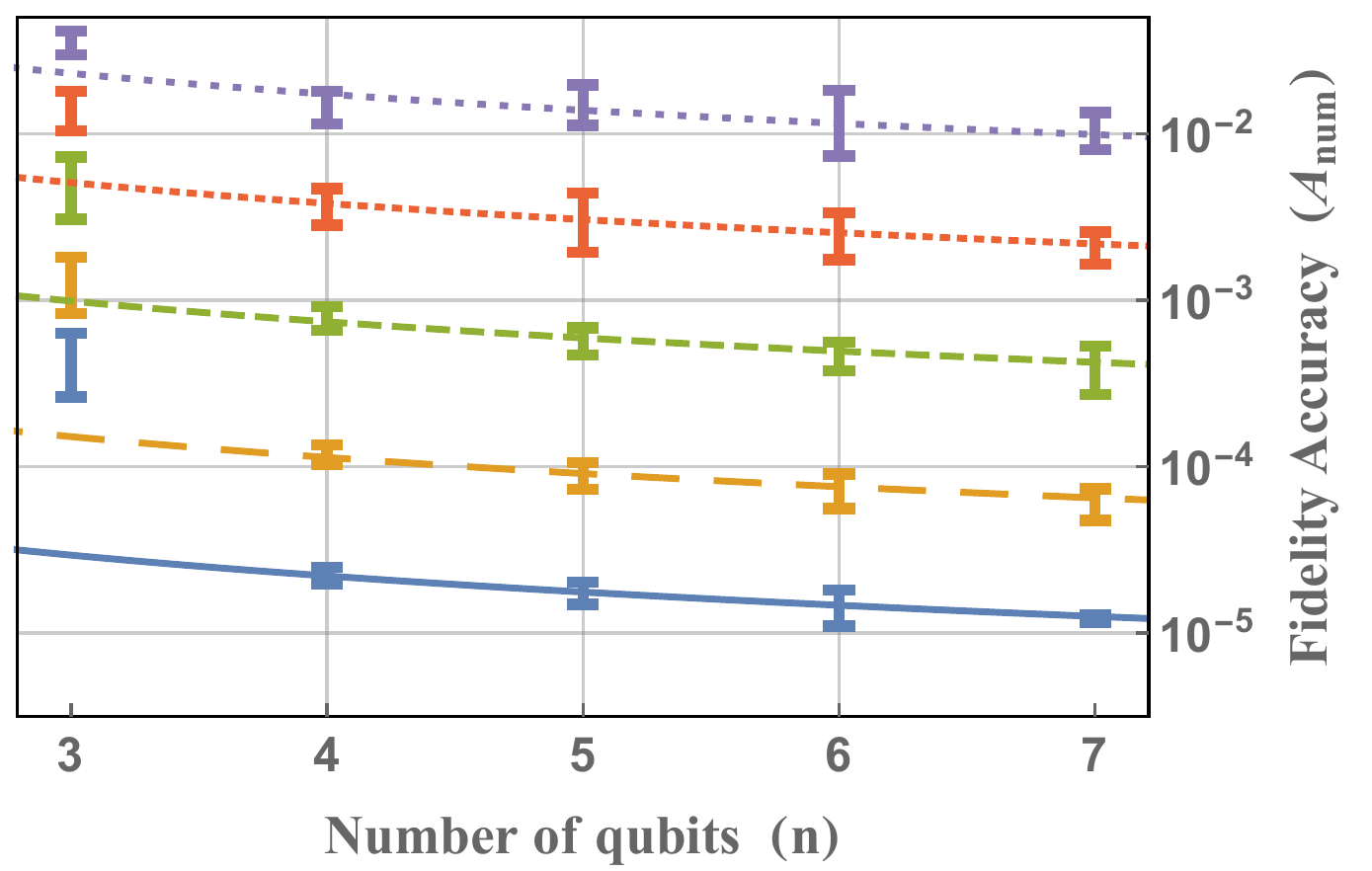}
\caption{\bf Numerical simulation of the experimental cost of finding a C-NOT gate in an Ising chain using steepest ascent \emph{in situ}.
\normalfont 
The number of updates and the fidelity accuracy required for the optimisation protocol to succeed is plotted for a chain of qubits with the Hamiltonian of \eqref{eq:systemHam} with nearest-neighbour Ising interactions, a gate time of $T_{\text{gate}}=4\pi$, and $N_{\text{ts}}=48$ timeslots. The target in each case is a CNOT gate between two qubits in the middle of the chain, separated by one other qubit. The gate fidelity used is $F_{LE}$, therefore the true gate fidelity will be a little higher. Since the cost of the $n=3$ case is significantly lower than for the other, it has been omitted from all fits.\\
Figure a) shows how $N_{\text{upds}}$ scales with the number of qubits for different target gate fidelities (error bars are twice the standard error). For this plot, the accuracy to which the fidelity is measured, $A_{\text{num}}$, is picked to give a $p_{\text{succ}} = 50\%$ success rate. We see a strong linear relation in the number of iterations required as a function of the number of qubits giving $N_{\text{upds}} = O(n)$.\\
Figure b) shows how the accuracy to which the local fidelity needs to be measured, $A_{\text{num}}$, scales with the number of qubits for different target fidelities (error bars are 5 times the standard error). The data is expected to have an $O(1/n)$ scaling as, in order to reach a gate infidelity of $\epsilon$, the fidelity ought to require a measurement accuracy $O(\epsilon)$.  As this is calculated from the sum of the fidelities of the subsystem, we conjectured that they each need to be measured to an accuracy $O(\epsilon/n)$. The data points lie very close to a $c/n$ curve, providing strong support for this argument. However, the constant $c$ does not appear to have quite a linear relationship with $\epsilon$; we did not investigate this further as it does not affect scalability.\\
The fidelity accuracy for the $p_{\text{succ}} = 50\%$ is estimated using an interpolation of $p_{\text{succ}}$ values for a range of $A_{\text{num}}$. Between 25 and 45 points are used in the interpolation. Each of these points are the average over a number of repetitions: 200 for $n=3,4$; 100 for $n=5$; 50 for $n=6,7$. The method for selecting the $A_{\text{num}}$ values for the simulations and the interpolations are described in more detail in the code repository \cite{qinsitu2018}.}
\label{fig:numIterPlot}
\end{figure*}

For the case of an Ising chain we also varied the number of qubits in order to hypothesise a likely scaling for $T_{\text{total}}$ in terms of the number of qubits in the system and found that a polynomial scaling matched very well. As the cost of doing a classical simulation of the \emph{in situ} optimisation of the Ising chain is much lower than for the others, we also picked this system to investigate the impact of measurement noise by introducing a finite value of $A_{\text{num}}$, which parametrises the sensitivity of the system to measurement noise. The results are shown in \figref{fig:numIterPlot} and give us a good estimate of $N_{\text{upds}}=O(n)$ and $A_{\text{num}}=O(1/n)$. The latter implies that $N_{\text{prec}} = O(n^2)$, due to the central limit theorem that states the number of repetitions required scales quadratically with the desired accuracy which gives $N_{\text{prec}} \propto A_{\text{num}}^{-2} = O(n^2)$.

Putting this together with the previous results that $N_{\text{fids}} = O(n)$ for gradient based optimisation and $N_{\text{meas}} = O(1)$ for the local estimator fidelity gives $N_{\text{runs}} = O(n^4)$. As this is done with a constant gate time and with a constant success probability, this implies that the time required to find a control sequence that implements a C-NOT gate on an Ising chain using a steepest-gradient \emph{in situ} scheme scales as $T_{\text{total}} = O(n^4)$.

\begin{figure*}
\centering
\includegraphics[scale=0.5]{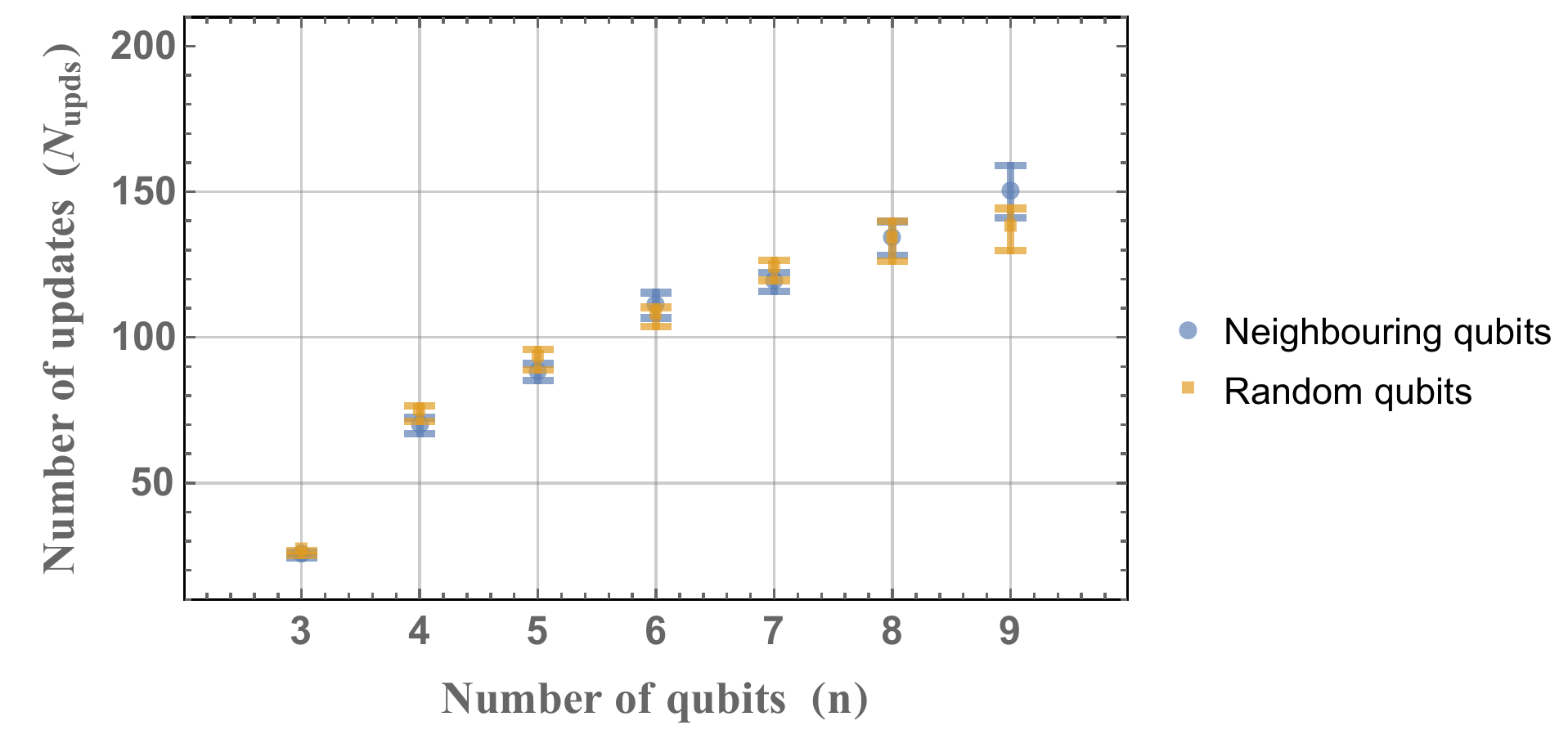}
\includegraphics[scale=0.5]{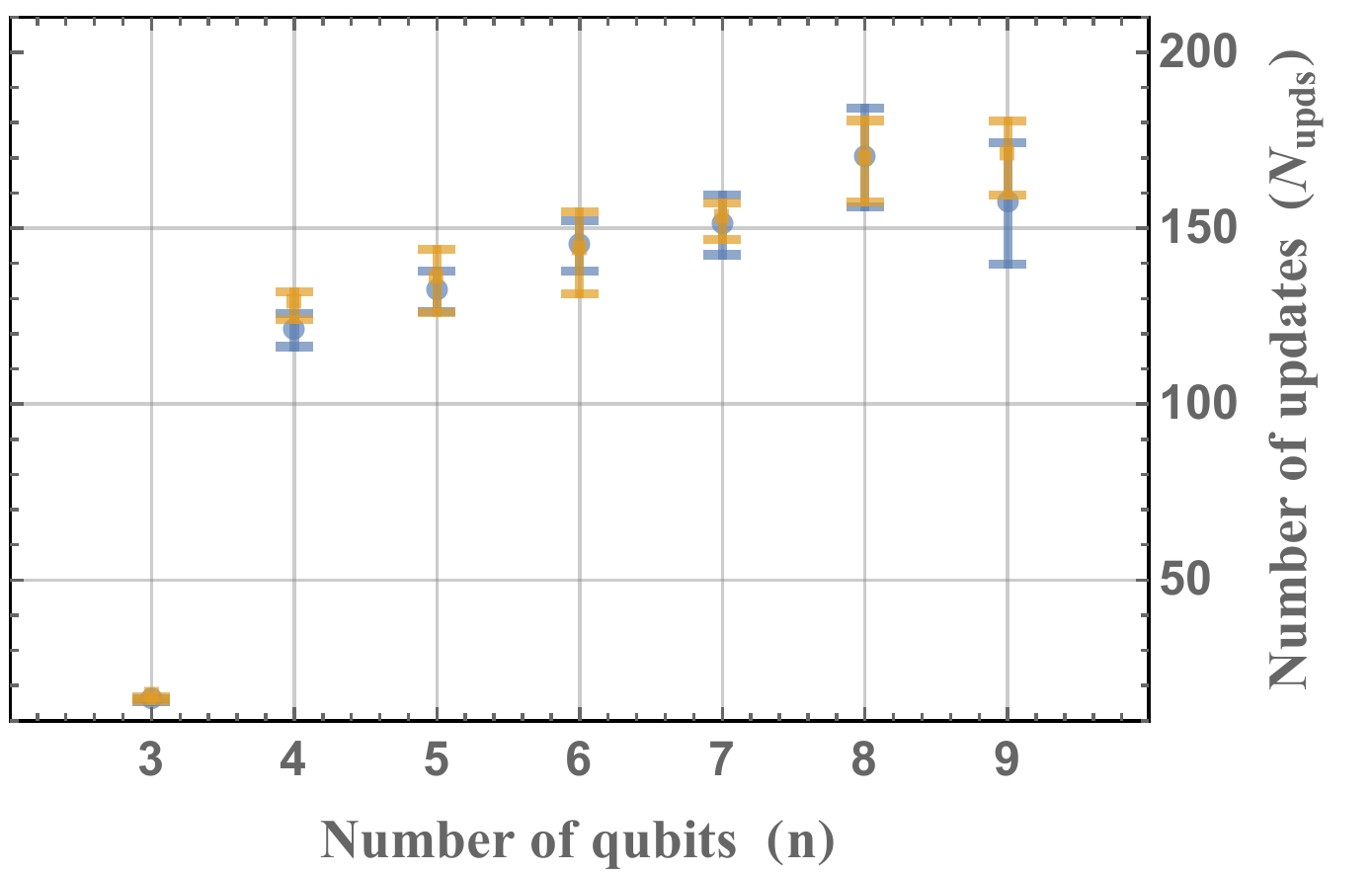}
\\
\small{a) Ising Chain \hspace{0.35\linewidth} b) Ising Ring} 
\caption{\bf Number of control pulse updates needed to find a C-NOT gate in an Ising chain and ring.
\normalfont 
The number of updates required for the optimisation protocol to succeed is plotted for a chain (left) and a ring (right) of qubits with the Hamiltonian of \eqref{eq:systemHam} with nearest-neighbour Ising interactions $\sigma_z\otimes\sigma_z$, a gate time of $T_{\text{gate}}=4\pi$, $N_{\text{ts}}=48$ timeslots, and a gate fidelity of $F_{LE} = 0.999$. Full numerical accuracy was used in these simulations. Each data point represents repeated optimisations: 100 for $n < 8$; 96 for $n = 8$; 30 for $n = 9$. The number of successful optimisations is $p_{\text{succ}} > 90\%$ in all cases. The error bars are twice the standard error.\\
In both cases there is no evidence that the scaling is more than linear, even disregarding the $n=3$ data. As we did not have an obvious model to fit to these points, no best-fit is shown. The value of $N_{\text{upds}}$ is slightly higher for the ring than the chain, but appears to have a smaller gradient in $n$. For both graphs the results for nearest-neighbour and random qubits are statistically indistinguishable. This highlights that the optimisation scheme operates equally well in both cases and works more efficiently than a naive dynamically decoupling protocol.}
\label{fig:ChainRing}
\end{figure*}

The other system we investigated in depth was an Ising ring where the target C-NOT was between two next-nearest-neighbour or two randomly located qubits. \figref{fig:ChainRing} shows the required $N_{\text{upds}}$ for up to $9$ qubits for this system. This graph is in agreement with the previous results of \figref{fig:numIterPlot} that the number of iterations required grows slowly with the number of qubits; in this case the results even show signs of being sub-linear. An unexpected feature shared by both sets of results is the low cost of the $3$ qubit case. Our intuition is that it is due to the additional symmetries present and the ease with which the single qubit not part of the target C-NOT gate can be kept disentangled.

The reason for picking the ring topology and not-nearest-neighbour target gate was to check whether the Ising chain results were unique, to demonstrate the applicability of the \emph{in situ} approach to different systems, and to test its ability to reach more complex gates. Specifically, it shows that the scalability of the scheme did not rely on boundary effects or on qubits being adjacent to each other. While a quantum computer could be built using only nearest-neighbour gates, being able to entangle two arbitrary qubits in the time of a single gate drastically reduces the potential run time of algorithms.

\subsection*{Discussion}

This polynomial scaling observed numerically in these two different cases is encouraging evidence that the protocol may indeed be efficient. Some of the components that make up this exponential scaling come from numerical data, so several fits are possible. However the points lie so close to a linear fit in \figref{fig:numIterPlot} that a different fit, such as an exponential one, would diverge only slowly. \figref{fig:ChainRing} suggests than corrections to the fit are more likely to make it sub-linear than more costly. Although it is clear that the results presented here do not form an absolute proof of the scalability of an \emph{in situ} control scheme for all quantum simulators, they are at the very least strong evidence that this is an powerful approach to take for  moderately large systems of a few tens of qubits.

Systems of such a size are interesting as they correspond to the state-of-the-art that can be realised experimentally. Using the \emph{in situ} scheme for such systems would likely find control sequences for entangling gates that are not currently known, and where purely classical numerical optimisation schemes would fail due to the enormous computational requirements. Furthermore, testing these predictions in such experiments would extend these results to numbers of qubits that are completely unattainable for a purely classical computer to model, and test this protocol closer to full-scale universal quantum computation.

One potential difficulty in optimal control is the existence of traps: local maxima of the fidelity that optimisation algorithms converge to which are not the global maxima. The question of whether traps exist in unitary control using the standard gate fidelity has been well studied \cite{Ho2009, Pechen2011, Russell2016}, and the conclusion is that generic quantum control landscapes are almost always trap free. This may also apply to the local estimator of the fidelity; traps were not a problem for the numerical simulations we performed and found no evidence of any new traps in \figref{fig:FidelityCurves} or elsewhere.

The numerical results presented here used GRAPE, which decomposed the control pulses into piece-wise constant functions. A potentially more powerful approach, and one which is harder to do classically but may be easier to implement physically, would be to decompose them by frequency \cite{Bartels2013}, such as in CRAB \cite{Doria2011, Caneva2011} and GOAT \cite{Machnes2015}. Such algorithms are slow to run classically due to the difficulty of exponentiating the time-dependent Hamiltonian, a step which is bypassed in the \emph{in situ} scheme. They typically require fewer parameters to describe a successful control pulse and thus may prove faster than GRAPE when done experimentally. A different variation would be to change from a gradient based algorithm to a geometric or genetic one, or even to use machine-learning algorithms to learn about the system \cite{Palittapongarnpim2016}. Ideas from robust control \cite{Daems2013, Barnes2015} may also be usable in an \emph{in situ} framework, in order to make the approach more resilient to fluctuations in the system Hamiltonian.

A future direction to take this work is to apply it to another important aspect of quantum computation: error correction. The protocol detailed here can be used in much the same way for this by replacing the target operation from a C-NOT gate, to one protecting some logical qubits. Preliminary results show that with a tuneable interaction the system can discover decoherence free subspaces and simple error correcting codes this way. Work remains on what the most useful tasks to optimise for are, and on showing the scalability of this approach.

\subsection*{Acknowledgements}

This work was supported by EPSRC through the Quantum Controlled Dynamics Centre for Doctoral Training, the EPSRC Grant No. EP/M01634X/1, and the ERC Project ODYCQUENT. We are grateful to HPC Wales for giving access to the cluster that was used to perform the numerical simulations. Many thanks to Stephen Glaser and David Leiner for discussions on possible implementations.

\small
\bibliographystyle{unsrtnat-etal}
\bibliography{library}
\normalsize

\clearpage
\section*{Appendix}
\label{sec:Methods}
\subsection*{Local estimator to the fidelity} We derive a bound for the gate fidelity $F(M,U) = \braket{\psi | \rho_M | \psi}$ which is the fidelity of the Choi states of the target gate $U$ with the realised channel $M$ using methods derived from \cite{Cramer2010}. A more in depth analysis is available in \cite{DiveThesis}. The Choi state of $U$ is defined as $\ket{\psi} = U \otimes \mathds{1} \ket{\Omega}$, and the Choi state of $M$ is $\rho = (M \otimes id) \ket{\Omega}\bra{\Omega}$, where $id$ is the identity map, and $\ket{\Omega} = \sum_k \ket{k k}$ is a maximally entangled state between the original Hilbert space and a copy of it. We consider the case where the target operation $U$ is unitary and has a tensor product structure such that $U = \bigotimes_i U_i$. The Choi state of $U$ inherits its tensor product structure and is given by $\ket{\psi} = \bigotimes_i \ket{\psi_i}$.

To find a bound for $F(M,U)$ we begin by constructing the projectors $h_i = \mathds{1}_i - \ket{\psi_i}\bra{\psi_i}$ for each $U_i$. These projectors have a very simple spectrum with a single $0$ eigenvalue with corresponding eigenket $\ket{\psi_i}$, and a degenerate orthogonal space with eigenvalue $1$. These projectors are summed together to form a Hamiltonian $H = \sum_i h_i \otimes \mathds{1}_{\bar{i}}$ such that each $h_i$ acts on its own part of the Hilbert space and is identity on the rest. This has a single $E_0 = 0$ eigenvalue with eigenstate $\ket{E_0} = \ket{\psi}$, while all its other eigenvalues are positive integers. By expanding this Hamiltonian in its eigenbasis $\{E_k, \ket{E_k}\}$ we have
\begin{align*}
\Tr{H \rho} &= \sum_{k\ge 0} E_k \braket{E_k|\rho|E_k} \\
&\ge \sum_{k>0} \braket{E_k|\rho|E_k},
\end{align*}
as $E_0 = 0$ and all the other energies are one or greater. Next, using the identity $\Tr{\rho}=\sum_{k\ge0} \braket{E_k | \rho | E_k} = 1$ we find
\begin{align*}
\Tr{H \rho} &\ge 1 - \braket{E_0|\rho|E_0}\\
&\ge 1 - \braket{\psi | \rho | \psi}.
\end{align*}
From the definition of the gate fidelity this give $F(M, U) \ge 1 - \Tr{H \rho}$.

The expectation value of the Hamiltonian can be evaluated according to
\begin{align*}
\Tr{H \rho} &= \sum_i \Tr{(\mathds{1}_i - \ket{\psi_i}\bra{\psi_i})\otimes\mathds{1}_{\bar{i}}\,\,\rho} \\
&= \sum_i \text{Tr}_i[(\mathds{1}_i - \ket{\psi_i}\bra{\psi_i}) \rho_i ] \\
&= \sum_i (1 - \braket{\psi_i | \rho_i | \psi_i}),
\end{align*}
where $\rho_i = \text{Tr}_{\bar{i}}[\rho]$. This is also the Choi state of the map
\begin{equation*}
M_i(\,\cdot\,) \equiv M(\,\cdot\, \bigotimes_{j\ne i} \tfrac{1}{d_j} \mathds{1}_j),
\end{equation*}
which is the map $M$ acting on subsystem $i$ with the other subsystems in the maximally mixed state. This results in \eqref{eq:fidbound}
\begin{equation*}
F(M, U) \ge 1 - \sum_i (1 - F(M_i, U_i)).
\end{equation*}

One way of measuring $F_{LE}$ is to prepare the state of one subspace in a basis state and all the others in a maximally mixed state, perform the simulation, measure the initial subspace, and then repeat for a tomographically complete basis set and for each subsystem; giving a cost of $N_{\text{meas}} = \sum_i O((d_i)^2) = O(n)$. However by noting that a maximally mixed state is a random mixture of pure states, the fidelity of each subsystem can be measured at the same time by preparing each one in a random pure basis state. In this case there is no scaling with the number of qubits as the number of repetitions required depends only on the size of the largest subsystem. This gives $N_{\text{meas}} = O(1)$.

While this shows that local fidelity is efficient to extract from an experimental set up, it is no easier than the gate fidelity to compute in numerical simulations. This is because it must be calculated from a unitary (or CPT map) which represents the evolution of the whole system of dimension $2^{n}$. Multiple partial traces of it are needed in order to compute the $M_i$ from \eqref{eq:fidbound}, which is typically a slow operation. Indeed, the need to calculate the local fidelity was a considerable strain on our numerical simulations and one of the compounding reasons why we could not simulate larger systems.

\end{document}